\documentclass[twocolumn,floatfix]{revtex4}
	\usepackage{amssymb}
	\usepackage{epsfig}
	\usepackage{amsmath}
	\usepackage{graphicx}
	
\begin{document}

\newcommand{\comment}[1]{}

\newcommand{\eps}{\epsilon}
\newcommand{\e}{{\rm e}}
\newcommand{\dm}{{\rm d}m}
\newcommand{\dx}{{\rm d}x}
\newcommand{\ds}{{\rm d}s}
\newcommand{\dt}{{\rm d}t}
\newcommand{\du}{{\rm d}u}
\newcommand{\ex}[1]{{\rm E}[#1]}
\newcommand{\poi}{{\rm Poisson}}

	\title{Why Mapping the Internet is Hard}
	\author{Aaron Clauset$^*$ and Cristopher Moore$^{*,\dagger}$}
	\affiliation{$^*$Computer Science Department and $^\dagger$Department of Physics and Astronomy, University of New Mexico, Albuquerque NM 87131 \\ 
	{\tt (aaron,moore)@cs.unm.edu}}
	\date{\today}

\begin{abstract}
Despite great effort spent measuring topological features of large networks like the Internet, it was recently argued that sampling based on taking paths through the network (e.g., traceroutes) introduces a fundamental bias in the observed degree distribution.  We examine this bias analytically and experimentally.  For classic random graphs with mean degree $c$, we show analytically that traceroute sampling gives an observed degree distribution $P(k) \sim k^{-1}$ for $k \lesssim c$, even though the underlying degree distribution is Poisson. For graphs whose degree distributions have power-law tails $P(k) \sim k^{-\alpha}$, the accuracy of traceroute sampling is highly sensitive to the population of low-degree vertices.  In particular, when the graph has a large excess (i.e., many more edges than vertices), traceroute sampling can significantly misestimate $\alpha$.
\end{abstract}
\maketitle

The Internet is a canonical complex network, and a great deal of effort has been spent measuring its topology.  However, unlike the Web, where the outgoing links 
 are directly visible, we cannot typically ask a router who are its neighbors. 
 As a result, studies have sought to {\em infer} the topology of the Internet by aggregating either paths through the network (i.e., traceroutes from a small number of sources to a large number of destinations)~\cite{Pansiot, Govindan, IMP, skitter, Opte}, routing decisions like those imbedded in BGP routing tables~\cite{BGP, Amini, Oregon}, or both~\cite{Faloutsos, Rocketfuel, LookingGlass}.  Although such methods are known to be noisy~\cite{Amini, Chen, Barford}, they strongly suggest that the Internet has a power-law degree distribution at both the router and domain levels.

However, Lakhina et al.~\cite{Lakhina} recently argued that traceroute-based sampling introduces a fundamental bias in topological inferences, since the probability that an edge appears within an efficient route decreases with the distance from the source. They showed empirically that traceroutes from a single source cause Erd\H{o}s-R\'enyi random graphs $G(n,p)$, whose underlying distribution is Poisson~\cite{gnp}, to appear to have a power law degree distribution $P(k) \sim k^{-1}$.

In this paper, we prove this result analytically by modeling the growth of a spanning tree on $G(n,p)$ using differential equations.  Certainly no one would argue that the Internet is a purely random graph; indeed, the degree distributions reported in e.g.~\cite{Faloutsos} have $P(k) \sim k^{-\alpha}$ with $2 < \alpha < 3$.  However, it is evocative that traceroute sampling can create the appearance of a power-law degree distribution where none in fact exists.

Even if the Internet has a power-law degree distribution, it is reasonable to ask whether traceroute sampling gives an accurate estimate of the exponent $\alpha$ (a question raised also in~\cite{DallAsta}). Here, we demonstrate that power-law degree distributions are only well sampled when the graph has a small {\em excess}, i.e., a mean degree close to $2$, so that the graph is very treelike. Other cases can result in a significant over- or under-estimation of $\alpha$.  Indeed, the accuracy of traceroute sampling is highly sensitive to the low-degree part of the degree distribution, not just the high-degree tail.

\bigskip

{\em Traceroute spanning trees: analytical results.}  
The set of traceroutes from a single source can be modeled as a spanning tree~\cite{mult}.  If we assume that Internet routing protocols approximate shortest paths, this spanning tree is built breadth-first from the source.  In fact, the results of this section apply to spanning trees built in a variety of ways, as we will see below.

We can think of the spanning tree as built step-by-step by an algorithm that explores the graph.  At each step, every vertex in the graph is labeled \mbox{{\em reached}}, \mbox{{\em pending}}, or \mbox{{\em unknown}}.  Pending vertices are the leaves of the current tree; reached vertices are interior vertices; and unknown vertices are those not yet connected.  We initialize the process by labeling the source vertex pending, and all other vertices unknown.  
Then the growth of the spanning tree is given by the following pseudocode:
\begin{quote}
\begin{tabbing}
{\tt while}\=\ there are pending vertices: \\
\>choose a pending vertex $v$ \\
\>label $v$ reached \\
\>for \=every unknown neighbor $u$ of $v$, \\
\> \> label $u$ pending.
\end{tabbing}
\end{quote}
The type of spanning tree is determined by how we choose the pending vertex $v$.  Storing vertices in a queue and taking them in FIFO (first-in, first-out) order gives a breadth-first tree of shortest paths; if we like we can break ties randomly between vertices of the same age in the queue, which is equivalent to adding a small noise term to the length of each edge as in~\cite{Lakhina}.  Storing pending vertices on a stack and taking them in LIFO (last-in, first-out) order builds a depth-first tree.  Finally, choosing $v$ uniformly at random from the pending vertices gives a ``random-first'' tree.

\begin{figure} [htbp]
\begin{center}
\includegraphics[scale=0.45]{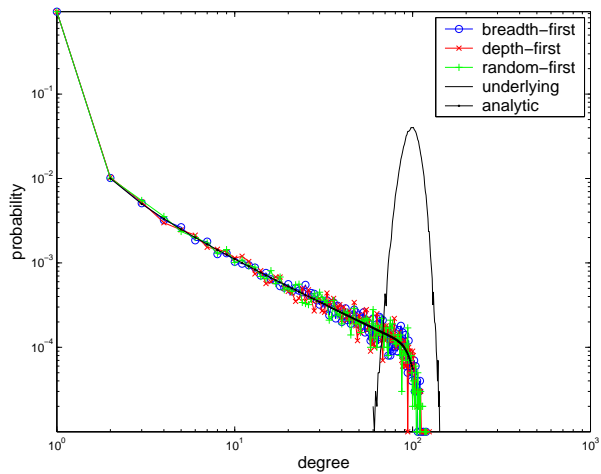}
\caption{Sampled degree distributions from breadth-first, depth-first and random-first spanning trees on a random graph of size $n=10^5$ and average degree $c=100$, and our analytic results (black dots). For comparison, the black line shows the Poisson degree distribution of the underlying graph. Note the power-law behavior of the apparent degree distribution $P(k) \sim k^{-1}$, which extends up to a cutoff at $k \sim c$. }
\label{fig:graph}
\end{center}
\end{figure}

Surprisingly, while these three processes build different trees, and traverse them in different orders, they all yield the same degree distribution when $n$ is large. To illustrate this, Fig.~\ref{fig:graph} shows the degree distributions for each type of \comment{breadth-first, depth-first, and random-first} spanning tree for a random graph $G(n,p=c/n)$ where $n=10^{5}$ and $c=100$.  The three degree distributions are indistinguishable, and all agree with the analytic results derived below.


We now show analytically that building spanning trees in Erd\H{o}s-R\'enyi random graphs $G(n,p=c/n)$ using any of the processes described above gives rise to an apparent power law degree distribution $P(k) \sim k^{-1}$ for $k \lesssim c$.  We focus here on the case where the average degree $c$ is large, but constant with respect to $n$; we believe our results also hold if $c$ is a moderately growing function of $n$, such as $\log n$ or $n^\eps$ for small $\eps$, but it seems more difficult to make our analysis rigorous in that case.


To model the progress of the {\tt while} loop described above, let $S(T)$ and $U(T)$ denote the number of pending and unknown vertices at step $T$ respectively.  The expected changes in these variables at each step are 
\begin{eqnarray}
\ex{U(T+1)-U(T)} & = & - p U(T) \nonumber \\
\ex{S(T+1)-S(T)} & = & p U(T) - 1 \label{eq:diff}
\end{eqnarray}
Here the $p U(T)$ terms come from the fact that a given unknown vertex $u$ is connected to the chosen pending vertex $v$ with probability $p$, in which case we change its label from unknown to pending; the $-1$ term comes from the fact that we also change $v$'s label from pending to reached.  Moreover, these equations apply no matter how we choose $v$; whether $v$ is the ``oldest'' vertex (breadth-first), the ``youngest'' one (depth-first), or a random one (random-first).  Since edges in $G(n,p)$ are independent, 
the events that $v$ is connected to each unknown vertex $u$ are independent and occur with probability $p$.  

Writing $t=T/n$, $s(t) = S(tn)/n$ and $u(tn)=U(t)/n$, the difference equations~\eqref{eq:diff} become the following system of differential equations,
\begin{equation}
\frac{\du}{\dt} = -c u \enspace , \quad
\frac{\ds}{\dt} = c u - 1 
\label{eq:sys}
\end{equation}
With the initial conditions $u(0) = 1$ and $s(0) = 0$, the solution to~\eqref{eq:sys} is
\begin{equation}
\label{eq:sol}
u(t) = \e^{-ct} , \quad s(t) = 1 - t - \e^{-ct} \enspace .
\end{equation}
The algorithm ends at the smallest positive root $t_f$ of $s(t) = 0$; using Lambert's function $W$, defined as $W(x) = y$ where $y \e^y = x$, we can write
\begin{equation}
\label{eq:t0}
 t_f = 1 + \frac{1}{c} W(-c\e^{-c}) \enspace .
\end{equation}
Note that $t_f$ is the fraction of vertices which are reached at the end of the process, and this is simply the size of the giant component of $G(n,c/n)$.

Now, we wish to calculate the degree distribution $P(k)$ of this tree.  The degree of each vertex $v$ is the number of its previously unknown neighbors, plus one for the edge by which it became attached (except for the root).  Now, if $v$ is chosen at time $t$, in the limit $n \to \infty$ the probability it has $k$ unknown neighbors is given by the Poisson distribution with mean $m = cu(t)$,
$\poi(m,k) = \e^{-m} m^k / k!$.
Averaging over all the vertices in the tree 
gives
\[ P(k+1) = \frac{1}{t_f} \int_0^{t_f} \dt \,\poi(cu(t),k) \enspace . \]
It is helpful to change the variable of integration to $m$.  Since $m = c \e^{-ct}$ we have $\dm = -cm \,\dt$, and 
\begin{eqnarray}
 P(k+1) & = & \frac{1}{t_f} \int_{c(1-t_f)}^c \dm \,\frac{\poi(m,k)}{c m} \nonumber \\
 & \approx & \int_{c \e^{-c}}^c \dm \,\frac{\poi(m,k)}{cm} \nonumber \\
 & = & \frac{1}{c k!} \int_{c \e^{-c}}^c \dm \,\e^{-m} m^{k-1} \enspace .
 \label{eq:p1}
\label{eq:int}
\end{eqnarray}
Here in the second line we use the fact that $t_f \approx 1 - \e^{-c}$ when $c$ is large (i.e., the giant component encompasses almost all of the graph).

The integral in~\eqref{eq:int} is given by the difference between two incomplete Gamma functions.  However, since the integrand is peaked at $m=k-1$ and falls off exponentially for larger $m$, for $k \lesssim c$ it coincides almost exactly with the full Gamma function $\Gamma(k)$.  Specifically, for any $c > 0$ we have
\[ \int_0^{c \e^{-c}} \dm \,\e^{-m} m^{k-1} < c \e^{-c} \]
and, if $k - 1 = c (1-\eps)$ for $\eps > 0$, then
\begin{eqnarray*}
 \int_c^\infty \dm \,\e^{-m} m^{k-1} 
 & = & \e^{-c} c^{k-1} \int_0^\infty \dx \,\e^{-x} (1+x/c)^{k-1} \\
 & < & \e^{-c} c^{k-1} \int_0^\infty \dx \,\e^{-x} \e^{x(k-1)/c} \\
 & = & \frac{\e^{-c} c^{k-1}}{\eps} 
 < \frac{\e^{-(k-1)} (k-1)^{k-1}}{\eps} \\
& \approx & \frac{\Gamma(k)}{\eps \sqrt{2 \pi (k-1)}}
 \end{eqnarray*}
\comment{
and, if $k - 1 = c - \Delta$ for $\Delta > 0$, then
\begin{eqnarray*}
 \int_c^\infty \dm \,\e^{-m} m^{k-1} 
 & = & \e^{-c} c^{k-1} \int_0^\infty \dx \,\e^{-x} (1+x/c)^{k-1} \\
 & < & \e^{-c} c^{k-1} \int_0^\infty \dx \,\e^{-x} \e^{x(k-1)/c} \\
 & = & \frac{\e^{-c} c^k}{\Delta} \\
 & < & \frac{c}{\Delta} \,\e^{k-1} (k-1)^{k-1} \\
 & \approx & \frac{c}{\Delta} \frac{\Gamma(k)}{\sqrt{2 \pi k}} \enspace .
\end{eqnarray*}
}
This is $o(\Gamma(k))$ if $\eps \gtrsim 1/\sqrt{k}$, i.e., if $k < c - c^\alpha$ for some $\alpha > 1/2$.  In that case we have
\begin{equation}
P(k+1) = (1-o(1)) \frac{\Gamma(k)}{c k!} 
\sim \frac{1}{ck}
\end{equation}
giving a power law $k^{-1}$ up to $k \sim c$.

Although we omit some technical details, this derivation can be made mathematically rigorous using results of Wormald~\cite{Wormald}, who showed that under fairly generic conditions, the state of discrete stochastic processes like this one is well-modeled by the corresponding rescaled differential equations.  Specifically it can be shown that if we condition on the initial source vertex being in the giant component, then with high probability, for all $t$ such that $0 < t < t_f$, $U(tn) = u(t)n + o(n)$ and $S(tn)=s(t)n+o(n)$.  It follows that with high probability our calculations give the correct degree distribution of the spanning tree within $o(1)$.


\bigskip

{\em Power-law degree distributions.}  
We now turn to the case where the Internet {\em does} have a power-law degree distribution $P(k) \sim k^{-\alpha}$, and demonstrate that unless the {\em excess}, i.e., the number of edges minus the number of vertices, is small, traceroute sampling can significantly misestimate $\alpha$.

There are several methods of constructing random graphs with power-law degree distributions and we use two to support our claim: the configuration model~\cite{configuration} in which the graph is random but conditioned on its degree distribution, and preferential attachment~\cite{pa} in which the graph is grown by a dynamical process and has a degree distribution with a power-law tail. 

\begin{figure} [htbp]
\begin{center}
\includegraphics[scale=0.45]{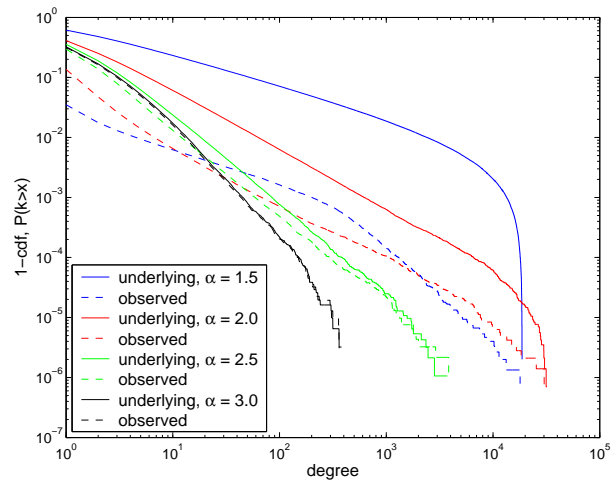}
\caption{Comparison of underlying and observed degree distributions in the configuration model with $n=5 \times 10^5$ and various $\alpha$.}
\label{fig:config}
\end{center}
\end{figure}

In the configuration model, we examined graphs where $P(k) = k^{-\alpha} / \zeta(\alpha)$ for all $k \ge 1$, with $\alpha$ ranging from $1.5$ to $3$.  Since these graphs are not necessarily connected, we compare the sampled and underlying degree distributions of the giant component (the latter has a power-law tail with the same exponent as the entire graph).
Fig.~\ref{fig:config} shows that as $\alpha$ increases, the observed distribution gets closer to the underlying distribution.  This closeness is because, for instance, when $\alpha = 3$ the ratio of edges to vertices in the giant component is only $1.02$ so its excess is only $0.02$ per vertex.  Thus {\em any} spanning tree on the giant component will include almost all of its edges, and sample its degree distribution fairly well.  

\begin{figure} [htbp]
\begin{center}
\includegraphics[scale=0.45]{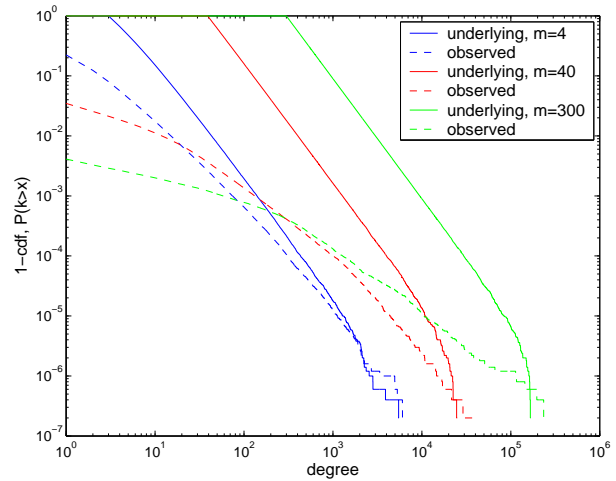}
\caption{Displacement of the power law tail for preferential attachment networks with $n=5 \times 10^5$. Traceroute sampling significantly underestimates the slope $\alpha$ as $m$ increases. }
\label{fig:pa}
\end{center}
\end{figure}

However, the size and excess of the giant component are highly sensitive to the low-degree part of the degree distribution, not just its power-law tail.  To illustrate this, Fig.~\ref{fig:pa} shows graphs grown using the preferential attachment model of~\cite{pa}.  Here every vertex has degree at least $m$, since it is given $m$ edges at birth. As $m$ increases, the slope of the observed distribution on a log-log plot becomes more shallow, giving a significant underestimate of $\alpha$; for instance, for $m=4$ we observe a slope of $2.7$ rather than the correct value $\alpha = 3$.  Using the configuration model to construct random graphs with a minimum degree $m$ and a degree distribution with a power-law tail yields similar results.

This underestimation of $\alpha$ occurs because traceroutes sample high-degree vertices more accurately than lower-degree ones: high-degree vertices are encountered early on in the breadth-first tree, when most of their neighbors are still unknown, while lower-degree vertices are encountered later, by which time most of their neighbors are already reached.  Thus the ``visibility'' of a vertex's edges increases with its degree~\cite{Lakhina}, making the slope of the observed distribution less negative.

\begin{figure} [htbp]
\begin{center}
\includegraphics[scale=0.45]{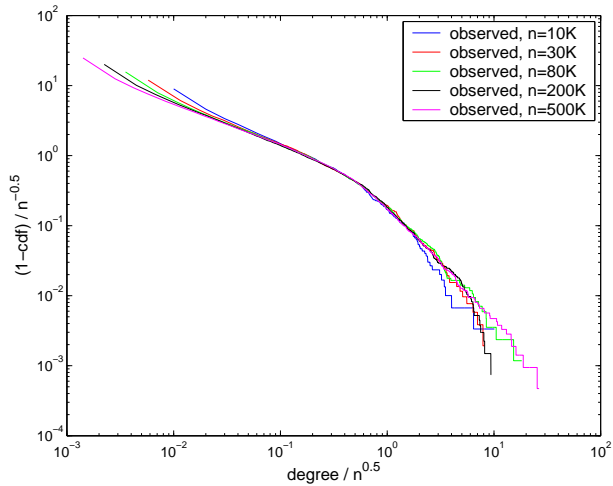}
\caption{Finite size effects for traceroute sampling with $\alpha=1.5$, with data collapse for various $n$.  Above a ``knee'' at $n^{0.5}$ the observed degree distribution falls off more sharply.}
\label{fig:collapsed}
\end{center}
\end{figure}

For smaller values of $\alpha$, Fig.~\ref{fig:config} shows that traceroute sampling encounters another kind of problem at smaller values of $\alpha$, namely significant finite-size effects.  The observed value of $\alpha$ is roughly correct up to a ``knee,'' above which the degree distribution falls off more sharply.  For $\alpha = 1.5$, for instance, Fig.~\ref{fig:collapsed} shows that this knee occurs at a degree $k \sim n^{0.5}$.  In these cases, a linear fit to the observed degree distribution will considerably over-estimate $\alpha$ unless we omit the data above the knee.

\bigskip

{\em Conclusions.}  There are two properties of the Internet which make it difficult to map: unlike the World Wide Web where links are visible, the Internet's topology must be queried indirectly, e.g., by traceroutes; and, since efficient routing protocols cause these traceroutes to approximate shortest paths, edges far from the source are difficult to see. It was observed by~\cite{Lakhina} that these effects can significantly bias the observed degree distribution, and even create the appearance of a power law where none exists.  We have proved this result analytically for random graphs $G(n,p)$, which yield an observed distribution $P(k) \sim k^{-1}$ for $k$ up to the average degree.
Other mechanisms by which power laws can appear in $G(n,p)$ include gradient-based flows ~\cite{jamming}, probabilistic pruning~\cite{paolo}, and minimum spanning trees on weighted random graphs~\cite{barabasi}.

While it seems likely that the Internet does have a power-law distribution, we have shown that traceroute sampling can signficantly misestimate the scaling exponent $\alpha$. Thus we suggest that the published values of $\alpha$ may not accurately reflect the real scaling of the Internet's topology.
This poses an interesting inverse problem: namely, given the value of $\alpha$ observed in traceroutes, what is the most likely value of $\alpha$ in the underlying graph? Also, since traceroutes from a single source, or a small number of sources (briefly explored in~\cite{Lakhina, DallAsta}), are inherently biased, how many sources are needed, as a function of network size and topology, to accurately sample the network?

We are grateful to David Kempe, Mark Newman, Mark Crovella, Paolo De Los Rios, Michel Morvan, Todd Underwood, Dimitris Achlioptas, Nick Hengartner and Tracy Conrad for helpful conversations. This work was funded by NSF grant PHY-0200909, and facilitated by Hewlett-Packard Gift No. 88425.1 to Darko Stefanovic.



\begin{thebibliography}{99}
\bibitem{BGP}
Autonomous Systems (ASs) use the Border Gateway Protocol (BGP) to route packets between networks. BGP routing tables represent logical domain adjacencies rather than physical connectivity. Tables contain advertised paths through this logical network; routing within domains is typically done via Open Shortest Path First or other distributed spanning tree protocol.
\bibitem{mult}
While there are typically multiple shortest paths from the source to a given destination, their union forms a tree if we break ties e.g.\ by giving each node an index and taking the shortest path which comes first in lexicographic order.
\end{thebibliography}
\end{document}